\documentclass[prd,aps,preprint,tightenlines,superscriptaddress]{revtex4}
\usepackage{epsfig}
\preprint{DOE/ER/40762-292} \preprint{UM-PP\#04-001}

\begin{document}
\title{Probing Quark Distribution Amplitudes Through \\ Generalized Parton Distributions at Large Momentum Transfer}
\author{Pervez Hoodbhoy}
\affiliation{Department of Physics, Quaid-e-Azam University,
Islamabad 45320, Pakistan} \affiliation{Institute for Nuclear
Theory, University of Washington, Seattle, WA 98975}
\author{Xiangdong Ji} \email{xji@physics.umd.edu} \affiliation{Department of
Physics, University of Maryland, College Park, Maryland 20742}
\affiliation{Institute for Nuclear Theory, University of
Washington, Seattle, WA 98975}\author{Feng Yuan}
\email{fyuan@physics.umd.edu} \affiliation{Department of Physics,
University of Maryland, College Park, Maryland 20742}
\affiliation{Institute for Nuclear Theory, University of
Washington, Seattle, WA 98975}
\date{\today}
\vspace{0.5in}
\begin{abstract}

In the large momentum transfer limit, generalized parton
distributions can be calculated through a QCD factorization
theorem which involves perturbatively-calculable hard kernels and
light-cone parton distribution amplitudes of hadrons. We
illustrate this through the $H_q(x,\xi,t)$ distribution for the
pion and proton, presenting the hard kernels at leading order. As
a result, experimental data on the generalized parton
distributions in this regime can be used to determine the
functional form of the parton distribution amplitudes which has
thus far been quite challenging to obtain. Our result can also be
used as a constraint in phenomenological GPD parametrizations.

\end{abstract}

\maketitle

It has been established in perturbative quantum chromodynamics
(pQCD) that hard exclusive processes in the asymptotic limit
depend on the non-perturbative light-cone parton distribution
amplitudes of hadrons\cite{exclusive,ex2,ex3,Duncan:hi}. However,
the functional form of these amplitudes in parton momenta $x_i$
has been very difficult to determine either from experimental data
or from theoretical calculations \cite{lattice, sumrule}. For
instance, the asymptotic electromagnetic form factors depend only
on an integral of the distribution amplitudes\cite{exclusive,ex3}.
In the pion-nucleus diffractive production of two jets, one can in
principle learn about the shape of the quark distribution
amplitude in the pion, but the process is not infrared safe as has
been pointed out recently \cite{dijet}. Poor knowledge on the
distribution amplitude in the proton has been the main obstacle in
deciding at what momentum transfer the asymptotic pQCD calculation
is relevant \cite{sterman,kroll,stefanis}.

In this paper, we study the generalized parton distributions
(GPDs) \cite{gpd,gpdreview} of hardons in the large momentum
transfer limit. The GPDs are a new class of hadron observables
which combine the physics of electromagnetic form factors and
Feynman parton distributions, and are related to quantum
phase-space distributions of the partons through Fourier
transformation \cite{wigner}. Apart from the renormalization
scale, they depend on the momentum transfer $t$ as in a form
factor, light-cone momentum $x$ as in a parton distribution, and
the projection of the momentum transfer along the light-cone
direction $\xi$, also known as the skewness parameter.

We report here that the GPDs in $-t\rightarrow \infty$ limit are
calculable through QCD factortization in which the
non-perturbative physics is included in the light-cone
distribution amplitudes of hadrons. Using this, the functional
form of the distribution amplitudes can be studied through the
GPDs' dependence on $x$ and $\xi$. Conversely, our result provide
a constraint on phenomenological GPD parametrizations. The GPDs at
large-$t$ can be measured, for example, from deeply-virtual
Compton scattering or hard exclusive meson production or
doubly-virtual Compton scattering in the kinematic regime $Q^2\gg
-t\gg \Lambda^2_{\rm QCD}$ in which the factorization theorems for
scattering amplitudes have been proven \cite{factorization}.
However, it can be experimentally challenging to measure the cross
sections in this regime because of additional power suppression in
$t$; we will not explore this issue here.

We illustrate our main point first by considering the generalized
parton distribution $H(x, \xi, t)$ for the pion, defined through
\begin{eqnarray}
H_q(x,\xi,t)&&\!\!= \!\frac{1}{2}\int \frac{d\lambda}{2\pi}
e^{i\lambda x} \left\langle
\pi;P'\left|\overline{\psi}_q\left(-\frac{\lambda}{2}n\right)\!\!
\not\! n \right.\right.   \nonumber\\
&&~~~~~\!\!\left.\left. \times {\cal
P}e^{-ig\int_{\lambda/2}^{-\lambda/2}d\alpha n\cdot A(\alpha n)}
\psi_q\left(\frac{\lambda}{2}n\right)\right|\pi;P\right\rangle \ ,
\label{pion}
\end{eqnarray}
where $P$ and $P'$ are the initial and final state pion momenta,
respectively, $t=(P-P')^2$, and ${\cal P}$ indicates path-ordering
for the light-cone gauge link. Introducing $\overline{P}=(P+P')/2$
along the $z$ direction and the conjugation light-cone four-vector
$n$, such that $n^2=0$ and $n\cdot \overline{P}=1$, the skewness
parameter is the projection of the momentum transfer $P'-P$ along
$\overline{P}$ direction, $\xi =-n\cdot (P'-P)/2$. The initial and
final light-cone momenta of the quarks are then $n\cdot k=x+\xi$
and $n\cdot k'= x-\xi$, respectively.

In the large momentum transfer limit, one can calculate the above
GPD using the pQCD factorization formalism which has been widely
applied to electromagnetic and other form factors
\cite{exclusive,ex2,ex3,f2}. The leading pQCD contribution is
shown in Fig. 1, where the initial and final pion states are
replaced by the light-cone Fock component with the minimal number
of partons. The circled crosses in the diagrams represent the
bilocal quark operator in Eq. (\ref{pion}). The hard part
responsible for the large momentum transfer contains a single
gluon exchange just like in the electromagnetic form factor. In
the first two diagrams (a) and (b), there is a hard gluon exchange
between the two quark lines, and in the third one, there is a
gluon coming from the gauge link. Since the transverse momenta of
the quarks are expected to be on the order of $\Lambda_{\rm QCD}$,
we may ignore them in calculating the hard part. Thus we can
effectively integrate out $k_\perp$ in the pion wave function to
obtain the distribution amplitude,
$\phi(x)=\int\frac{d^2k_\perp}{(2\pi)^3}\psi(x,k_\perp)$. The
parton transverse momenta flowing into the hard part are now taken
to be zero.

\begin{figure}[t]
\begin{center}
\includegraphics[height=2.5cm]{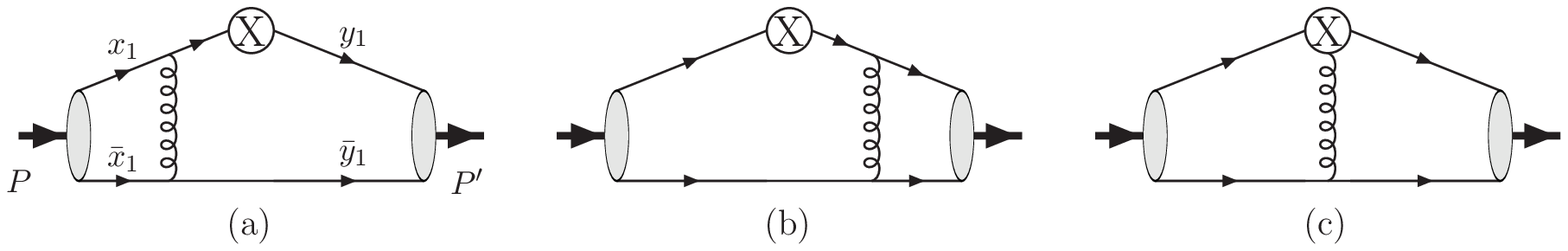}
\end{center}
\caption{\it Leading pQCD diagrams contributing to the pion's
generalized parton distribution $H(x, \xi, t)$ at large $-t$. The
circled crosses represent the non-local quark operator.}
\end{figure}

The result of the above pQCD analysis is a factorization formula
for the GPD at large $t$ in terms of the quark distribution
amplitude
\begin{eqnarray}
H_q(x,\xi,t, \mu) =\int dx_1 dy_1\phi^*(y_1, \mu)\phi(x_1, \mu)
T_{Hq}(x_1,y_1, x, \xi, t, \mu) \ ,
\end{eqnarray}
where $T_{Hq}$ is the hard part and can be calculated as a
perturbation series in $\alpha_s$. All quantities in the above
equation depend on the renormalization scale $\mu$. The
$\mu$-dependence in the hard part must be such that it accounts
for the difference between the GPD and the distribution amplitude.

The leading contribution to the hard part can be calculated
straightforwardly,
\begin{eqnarray}
T_u(x,x_1,y_1) &=& \frac{4\pi
\alpha_sC_F}{\overline{x}_1\overline{y}_1 (-t)}
\delta(x-\lambda_1)\left[(1-\xi) +
\frac{1-\xi^2}{\lambda_1-\tilde\lambda_1} \right]
  +{\rm h.c.}\ ,
\end{eqnarray}
where $C_F=4/3$, h.c. stands for a term obtained by exchange $x_i$
and $y_i$, and $\xi$ and $-\xi$, $\lambda_1=y_1+\overline y_1\xi$,
$\tilde\lambda_1=x_1-\overline x_1\xi$ ($\bar x= 1-x$).
 Since $0<y_1<1$, the first term
contributes when $x>\xi$; whereas the second term contributes when
$x>-\xi$, which indicates an up-anti-up pair contribution. The
anti-quark is generated through the one-gluon exchange on the top
of the valence wave function. The GPD for the down quark $H_d(x,
\xi, t)$ can be obtained from that of the up quark through simple
charge symmetry, $H_d(x,\xi,t) = -H_u(-x, \xi, t)$.

The above result can be translated into one for the moments of the
GPDs $H_q^{(n)}(\xi,t)=\int^1_{-1} dx x^{n-1}H_q(x,\xi,t)$. In
fact, the factorization formula applies for the individual
moments, $ H_q^{(n)}(\xi,t)=\int dx_1 dy_1\phi^*(y_1)\phi(x_1)
T_{Hq}^{(n)}(x_1,y_1, \xi,t),$ where $T_{Hq}^{(n)}(x_1,y_1,
\xi,t)$ is simply the $n$-th moment of $T_{Hq}(x_1,y_1, x,\xi,t)$.
For the up quark in the pion, we have,
\begin{eqnarray}
T_u^{(n)}(\xi,t)&=&\frac{4\pi\alpha_sC_F}{\overline{x}_1\overline{y}_1
(-t)} \left[(1-\xi)\lambda_1^{n-1}+
      (1+\xi)\tilde{\lambda}_1^{n-1} + (1-\xi^2)\sum_{m=0}^{n-2}\lambda_1^{m}
      \tilde{\lambda}_1^{n-m-2}\right]\ ,
\end{eqnarray}
which contains both even and odd powers of $\xi$. For $n=1$, the
above reproduces the hard part in the QCD factorization formula
for the pion form factor \cite{exclusive,ex2,ex3}. For $n=2$, $
T_u^{(2)}(\xi,t)=4\pi\alpha_sC_F/(\overline{x}_1\overline{y}_1
t)\left[(x_1+y_1+1)+2(x_1-y_1)\xi + (x_1+y_1-3)\xi^2 \right]$. It
is easy to see that the linear-dependence in $\xi$ does not
contribute to $H_u^{(1)}(\xi, t)$ because of the symmetry in the
initial and final states. For the same reason, all odd powers of
$\xi$ in $T^{(n)}_u$ do not contribute to the GPD moments.

\begin{figure}[t]
\begin{center}
\includegraphics[height=5.5cm]{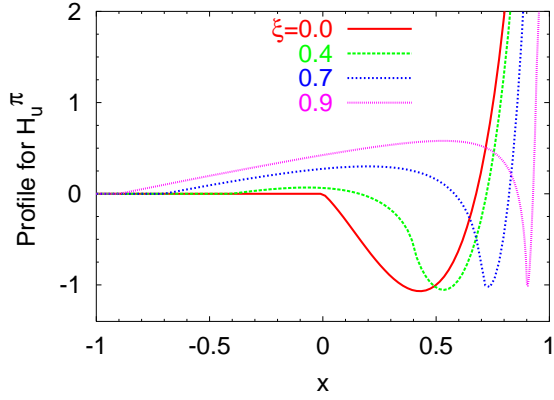}
\end{center}
\caption{\it The profile function of $H(x, \xi, t)$ for the pion
in the asymptotic limit.}
\end{figure}

It has been suggested from the dijet production \cite{dijet} and
$\gamma^*\gamma\rightarrow \pi$ transition \cite{transition} that
the pion distribution amplitude at $\mu>2$ GeV is very close to
the asymptotic amplitude $\sqrt{6}f_\pi x(1-x)$
\cite{exclusive,ex2,ex3}. If so, we can make the prediction for
the $H_u$ as follows,
\begin{eqnarray}
     H_u(x, \xi, t) &=& \frac{16\pi
\alpha_sf_\pi^2}{(-t)} \left\{ \theta(x-\xi)
  \frac{(x-\xi)}{(1-\xi)} \right. \\
&&\left.\times\left[-1-\frac{(x+\xi)}{(1+\xi)}\log\frac{(1-x)^2}{(x+\xi)^2}\right]
  + (\xi\rightarrow -\xi)  \right\}\nonumber\ ,
\end{eqnarray}
which is continuous at $x=\xi$ and $x=-\xi$. The quantity in the
braces is a profile function and is plotted for four different
$\xi$ in Fig. 2. The function diverges at $x=1$, indicating the
breaking down of $1/t$ expansion. This divergence generates a slow
decrease of the GPD moments at large $n$, and is present even when
$\xi=0$. We note that the limit $x->1$ and $-t\rightarrow \infty$
may not be interchangeable. If we take the limit $x\rightarrow 1$
first, $H_q$ may not vanish in the subsequent $-t\rightarrow
\infty$ limit because the pion momentum is now carried by a single
quark.

 Now we turn to the proton case. The factorization
formula for the GPD $H_q$ takes a similar form
\begin{eqnarray}
H_q(x,\xi,t)&=&\int
[dx][dy]\Phi_3^*(y_1,y_2,y_3)\Phi_3(x_1,x_2,x_3) T_{Hq}(x_i,y_i,
x, \xi, t) \ ,
\end{eqnarray}
where $[dx]=dx_1 dx_2dx_3\delta(1-x_1-x_2-x_3)$, and $\Phi_3(x_i)$
is the three-quark distribution amplitude \cite{braun}.

In the leading order in $\alpha_s$, there are three classes of
diagrams contributing to the hard part, each with a representative
shown in Fig. 2. The first class consists of diagrams with
two-gluon exchanges not attached to the non-local operators. Shown
in Fig. 2.1 is one of the 14 possible diagrams in this class. The
second class has one-gluon coming from the gauge link in the
non-local operator, and the third class with two gluons. To
calculate the hard part, we arrange the first quark to have spin
up, the second spin down, and the third spin up again, with
Feynman momentum $x_1$, $x_2$, $x_3$ for the incoming quarks, and
$y_1$, $y_2$, and $y_3$ for outgoing quarks, respectively. We use
$T_i$ to denote the hard part with the non-local operator inserted
on the line $i$. Then the hard part from the proton is
\begin{eqnarray}
T_u^p &=& \frac{1}{3}(2T_1 + T_2 + T_3+  T_1'+T_3') \nonumber \\
T_d^p &=& \frac{1}{3}(T_2+T_3+T_2')
\end{eqnarray}
$T'_i$ is obtained from $T_i$ by interchanging $y_1$ and $y_3$.

\begin{figure}[t]
\begin{center}
\includegraphics[height=2.5cm]{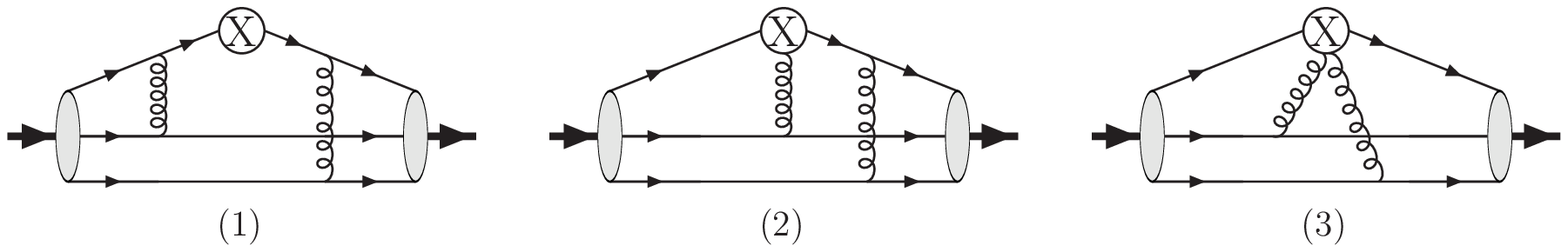}
\end{center}
\caption{\it Representatives from three classes of QCD diagrams
contributing to the proton GPD $H_q(x, \xi, t)$ in the asymptotic
limit. }
\end{figure}

Our result for the hard part is
\begin{eqnarray}
T_1&=&  \frac{2\pi^2C_B^2\alpha_s^2}{t^2}\left\{K_{11}
      \delta(x-\lambda_1)+K_{12}\frac{\delta(x-\lambda_1)-\delta(x-\tilde{\lambda}_1)}
      {\lambda_1-\tilde\lambda_1}\right.\nonumber\\
  &&+ \left.
      K_{13}\frac{\delta(x-\lambda_1)-\delta(x-\eta_1)}
      {\lambda_1-\eta_1} +K_{14}\frac{\delta(x-\lambda_1)-\delta(x-\eta_1)}
      {(\lambda_1-\eta_1)(\eta_1-\tilde\lambda_1)}\right\}+
      {\rm h.c.}\\
T_2
&=&\frac{2\pi^2C_B^2\alpha_s^2}{t^2}\left\{K_{21}\delta(x-\eta_2)
+\left[K_{13}'\frac{\delta(x-\lambda_2)-\delta(x-\eta_2)}{\lambda_2-\eta_2}
+(1\leftrightarrow 3)\right]
\right.\nonumber\\
&& \left. +K_{14}'\frac{\delta(x-\lambda_2)-\delta(x-\eta_2)}
      {(\lambda_2-\eta_2)(\eta_2-\tilde\lambda_2)}\right\}+
      {\rm h. c.}\ ,
\end{eqnarray}
where $C_B=2/3$, $\lambda_i=y_i+\overline y_i\xi$,
$\eta_1=1-x_3-y_2+(y_2-x_3)\xi$, and $\eta_2=
\eta_1(1\leftrightarrow 2)$. The functions $K_{ij}$ are defined as
\begin{eqnarray}
K_{11}&=&      \frac{1}{x_3y_3\overline x_1^2\overline y_1^2}+
      \frac{1}{x_2y_2\overline x_1^2\overline y_1^2}-
      \frac{1}{x_2y_2x_3y_3\overline{x}_3\overline y_1}\nonumber\\
K_{12}&=&\frac{1-\xi}{x_2y_2\overline x_1^2\overline y_1}
       +\frac{1-\xi}{x_3y_3\overline x_1^2\overline y_1}\nonumber\\
K_{13}&=&\frac{1-\xi}
    {x_2y_2x_3y_3\overline x_3}\nonumber\\
K_{14}&=&\frac{(1+\xi)(1-\xi)}{x_2y_2x_3y_3}\nonumber\\
K_{21}&=&\frac{1}{x_1y_1x_3y_3\overline x_3\overline y_1}\ ,
\end{eqnarray}
and $K_{ij}'=K_{ij}(1\leftrightarrow 2)$. $T_3$ can be obtained
from $T_1$ by exchanging $(x_1,y_1)$ and $(x_3, y_3)$,
respectively. From the above, we can calculate the GPD moments for
the nucleon in a factorization form. If we take the first moment,
we recover the pQCD prediction for the Dirac form factor
$F_1(Q^2)$ \cite{exclusive,ex2,ex3}. If we take the second moment,
we find the pQCD prediction for gravitational form factors
$A(Q^2)$ and $C(Q^2)$ at large $Q^2$ \cite{gpd}. The contribution
to $C(Q^2)$ is zero at this order in $1/t$. This is because
$C(Q^2)$ also contribute to the helicity-flip GPD $E(x,\xi,t)$
which is sub-leading in the large $-t$ limit.

\begin{figure}[t]
\begin{center}
\includegraphics[height=5.5cm]{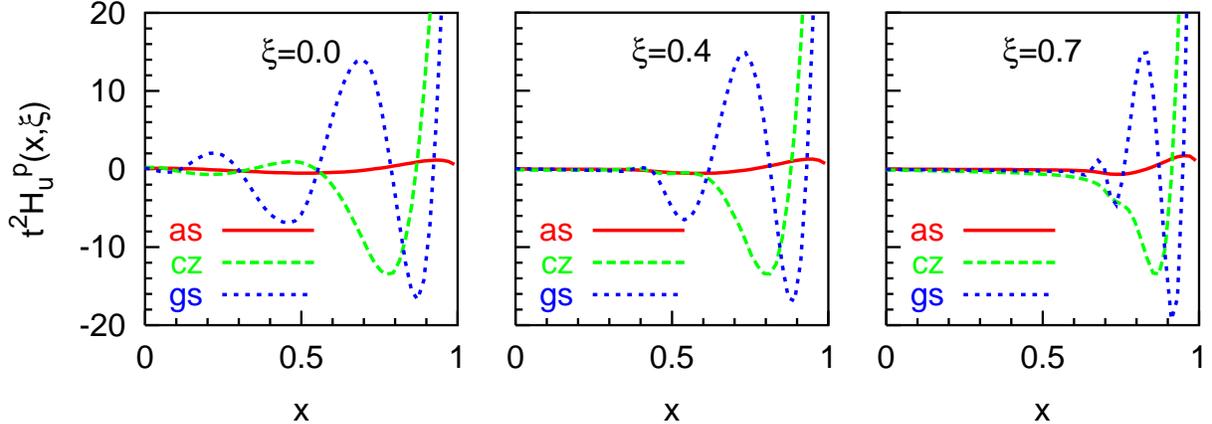}
\end{center}
\caption{\it $t^2H_u(x, \xi, t)$ for the proton at $-t =$ 20
GeV$^2$. CZ refers to the Chernyak and Zhitnitsky amplitude, GS
the Gari and Stefanis amplitude, and AS the asymptotic amplitude.}
\end{figure}

One can make a numerical calculation of $H_u(x,\xi,t)$ using
various model amplitudes in the literature
\cite{ex3,lattice,stefanis}. Using the strategy of Ref.
\cite{jisill}, we have computed $t^2H_u$, shown in Fig. 4, for 3
different $\xi$ at $t^2= -20$ GeV$^2$ with the asymptotic,
Chernyak-Zhitnitsky, and Gari-Stefanis amplitudes. Although the CZ
and GS amplitudes both give reasonable account of data on $F_1^p$
for $-t\ge 10$ GeV$^2$, the two yield very different predictions
for the GPD. Note that the scale of $H_u$ is strikingly large; a
relatively small Dirac $F_1$ is resulted from the cancellation in
the integration.

In summary, we have obtained a QCD factorization formula for the
generalized parton distributions in terms of the non-perturbative
light-cone distribution amplitudes and perturbatively-calculable
hard kernels. We have calculated the hard kernels for the pion and
the nucleon in the leading-order in $\alpha_s$. As a result, data
on the GPDs in the large-$t$ regime provides a way to constrain
the functional form of the distribution amplitudes.

Note added: After this paper was finished, we learned that the
pion case has been studied in Ref. \cite{new1}, our result differs
from that in the paper. Studying GPD in the large-t limit was
first done in \cite{new2} for $\gamma^*\gamma\rightarrow \pi\pi$.
We thank Diehl for pointing this out to us.

We are grateful for useful conversations with Stan Brodsky,
Jian-Ping Ma, and Dieter M\"uller. This work was supported by the
U. S. Department of Energy via grant DE-FG02-93ER-40762.

\end{document}